\documentclass[a4paper,11pt]{article}
\usepackage{pos}

\title{Model independent search for transient multimessenger events with AMON using outlier detection methods}
\ShortTitle{Multimessenger search for transient with AMON using outlier detection methods}

\manuallySeparateAuthors{}

\author*[a]{T. Grégoire,}
\author[a]{ H. A. Ayala Solares,}
\author[a]{ S. Coutu,}
\author[a]{ D. Cowen,}
\author[a]{ J. J. DeLaunay,}
\author[a]{ D. B. Fox,}
\author[b]{ A. Keivani,}
\author[a]{ F. Krauss,}
\author[a]{ M. Mostafá,}
\author[a]{ K. Murase,}
\author[a]{ E. Neights}
\author[a]{ and C. F. Turley}
\author{ for the AMON group}

\affiliation[a]{Pennsylvania State University, Department of Physics\\
  State College, USA}
\affiliation[b]{Columbia University, Department of Physics,\\
New York, USA}

\emailAdd{tmg5746@psu.edu}
\emailAdd{hza53@psu.edu}
\emailAdd{sxc56@psu.edu}
\emailAdd{jjd330@psu.edu}
\emailAdd{dbf11@psu.edu}
\emailAdd{azadeh.keivani@columbia.edu}
\emailAdd{felicia.krauss@psu.edu}
\emailAdd{mam1264@psu.edu}
\emailAdd{kum26@psu.edu}
\emailAdd{cft114@psu.edu}

\renewcommand{\printHeadAuthors}{T. Grégoire} 

\abstract{The Astrophysical Multimessenger Observatory Network (AMON) receives subthreshold data from multiple observatories in order to look for coincidences. Combining more than two datasets at the same time is challenging because of the range of possible signals (time windows, energies, number of events…). However, outlier detection methods can circumvent this issue by identifying any signal divergent from the background (e.g.\ scrambled data).

We propose to use these methods to make a model independent combination of the subthreshold data of neutrino and gamma ray experiments. Using the python outlier detection (PyOD) package, it allows us to test several methods from a simple ``k-nearest neighbours'' algorithm to a more sophisticated Generative Adversarial Active Learning neural networks which generates data points to better discriminate inliers from outliers.}

\FullConference{37$^{\rm{th}}$ International Cosmic Ray Conference (ICRC 2021)\\
		July 12th -- 23rd, 2021\\
		Online -- Berlin, Germany}

\begin{document}
\maketitle

\section{AMON}
The last decades have seen the emergence of multimessenger astrophysics. Indeed, the universe is now studied through the observation of photons, cosmic rays, neutrinos as well as gravitational waves and the combined observations of a source from multiple messengers has proven to be enlightening. The coincident detection of gravitational waves and electromagnetic radiations allowed the first detection of the coalescence of a binary neutron star~\cite{GW_GRB_170817}. Multiple messengers bringing different information are very instructive when put together. That is also the case of the first evidence of a high energy neutrino source~\cite{TXS_IC170922A, IC_170922A} from the coincident detection of neutrinos and a gamma ray flare.

The Astrophysical Multimessenger Observatory Network (AMON)~\cite{AMON} has been developed at the Pennsylvania State University, with the goal to combine subthreshold data from several astrophysical observatories near realtime. Indeed, AMON has signed Memoranda of Understanding (MoU) with different collaborations in order to receive their data below the discovery threshold in real-time.
Currently IceCube and ANTARES send subthreshold track events to AMON\@. We also receive IceCube high energy events above the detection threshold, the ``Gold'' and ``Bronze'' track events~\cite{IC_Gold_Bronze} as well as the cascades~\cite{IC_Casc}. HAWC sends two real-time datasets to AMON, the ``hotspots'' and the ``bursts''~\cite{HAWC_Bursts}. Fermi-LAT data are also stored in the AMON database. All the data received are stored in the AMON servers in order to do archival analyses.

The subthreshold data are background dominated and cannot be used to identify a signal alone, however if an excess is detected by multiple instruments their combined signal can become significant. The AMON team is looking for such signal by analysing the data in real-time thanks to the AMON infrastructure. AMON sends any statistically significant result publicly to the Gamma-ray Coordinate Network (GCN) so that small field-of-view instruments can point toward the direction of the signal, looking for a counterpart.
The data received are also stored in the AMON servers in order to do archival analyses.

By doing so, AMON contributes to the search and study of the most energetic phenomena in the universe, helping its partners to best exploit their data in order to answer fundamental questions of astrophysics, fundamental physics and cosmology.

The analysis presented in this proceeding aims at combining more than two datasets at the same time, which is something AMON was specifically designed to do.

\section{Multiple Datasets Outlier Detection}
We present here a search for coincident signal in several datasets based on outlier detection methods. Indeed, the search combining several datasets must allow the detection of a large range of signal coincidences, and it would not be feasible to simulate realistically all possible signal combinations and quantify the probability of one combination in respect to an other one. Outlier detection methods permit an agnostic search for signal by learning the background in order to classify any divergent data point as signal. In contrast to simulating signal, simulating background by scrambling the data is a straight forward task.

The method presented here is mostly independent from the datasets used as inputs. It could be applied to data currently available to AMON as well as future new data. We present the method in the form of an archival search, but it could also be used for a real-time stream in the future.

\subsection{Input Data}
\begin{figure}
  \centering
  \includegraphics[scale=0.45]{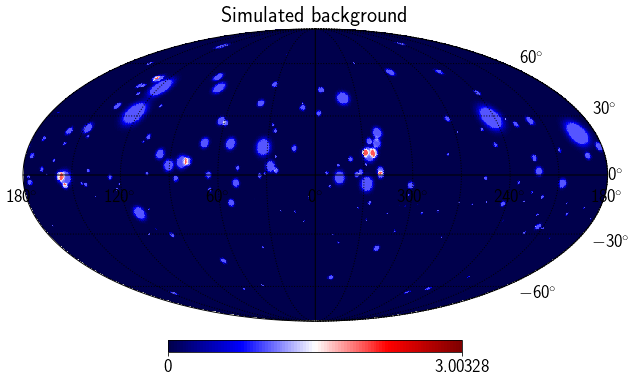}
  \parbox{5in}{\hspace*{1.5em}\caption{Skymap of the event density of a simulated background for a dataset with a large range of angular error sizes, for illustration purpose.}\label{fig:skymap}}
\end{figure}

This analysis takes as input a list of events with their corresponding date, position and position uncertainty for each dataset. These data are converted into skymaps of event densities for each time steps of 6h as illustrated in Fig.~\ref{fig:skymap}. To avoid that two events close in time fall within different time windows, the skymaps are done twice with a 3h shift in time.

\begin{figure}
  \centering
  \includegraphics[scale=0.45]{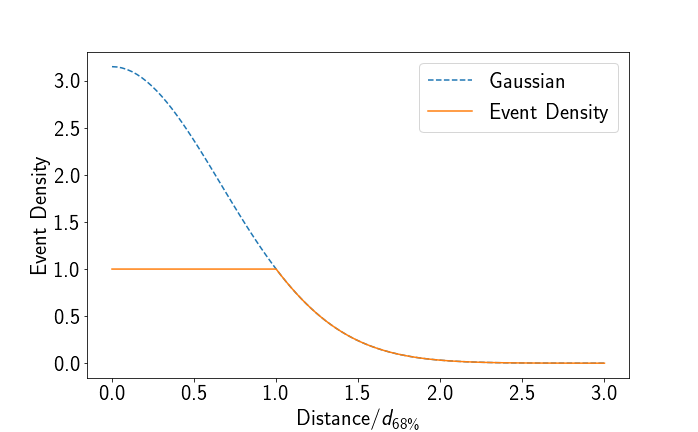}
  \parbox{5in}{\hspace*{1.5em}\caption{Event density of an event in a pixel as a function of the distance.}\label{fig:event_density}}
\end{figure}

The event density is defined as being unity for pixels within the 68\% error region of the event and it decreases following a 2D Gaussian for larger distances, as shown in Fig.~\ref{fig:event_density}. The Gaussian is scaled to be unity at the 68\% error contour $d_{68\%}$ for continuity, for a 2D Gaussian $d_{68\%} \approx 1.515\sigma$.

\begin{equation*}
  \text{event density} = \left\{
    \begin{array}{ll}
      1, & \text{ if } d < d_{68\%}\\
      \exp\left(\displaystyle\frac{-d^2 + d_{68\%}^2}{2\sigma^2}\right), & \text{otherwise}
    \end{array}
  \right.
\end{equation*}

This event density is chosen as it gives a larger value in the case of a pixel surrounded by a few events (e.g., an event density =\,3 for 3 nearby events) than in the case of only one centred event (=\,1) while the use of a Gaussian would not always allow one to distinguish a pixel with several nearby events from a pixel containing only one centered event. However, we plan to test different event density definitions.

The skymaps of the event densities are used to build the input data of the outlier detection algorithm. Each data point corresponds to a pixel of a time step and contains $n$ event densities corresponding to the $n$ datasets to combine, as well as the altitude and azimuth of the pixel seen from the so-called ``Null Island'' corresponding to the $0^\circ$\,N, $0^\circ$\,E in Earth coordinates.

\subsection{Scrambling}
The training of the outlier detection algorithm should be done on background only, therefore a scrambling of the data is done to simulate background. The scrambled data are also used to blind the analysis. The scrambling is typically done by permuting the azimuth and time of the events of a dataset. The equatorial coordinates are then computed from the scrambled local coordinates.

\subsection{Outlier Detection Algorithms}
Several outlier detection algorithms exist, therefore the PyOD (\textit{Python Outlier Detection})~\cite{PyOD} library was used as it implements many algorithm. We will test several of them and choose the one that fits best to our needs. Here is a short description of some of these algorithms:
\begin{itemize}
    \item \textit{K-nearest neighbours (KNN)}~\cite{KNN1, KNN2}: One of the simplest algorithm but still effective. The outlier score of a data point is its distance to its k$^\text{th}$ nearest neighbour.
    \item \textit{Histogram-based Outlier Score (HBOS)}~\cite{HBOS}: Very fast algorithm but less precise as it assumes the independence of the features.
    \item \textit{Principal Component Analysis (PCA)}~\cite{PCA}: Consists in a decomposition of correlated variables into a lower dimensional space of uncorrelated variables, the so-called ``principal components''. The outlier score of a data point is the sum of its projected distances on the principal components. However, in our case the input data are made of a few uncorrelated variables as the background of the different datasets are independent, while PCA is best for cases when the input is made of many partially correlated variables.
    \item \textit{AutoEncoder}~\cite{outlier_analysis}: A neural network that learns to encode a set of data into a lower dimension space and decode it back to its initial values. The error between the input and the output should be small only if the input is similar to the training sample, therefore the outlier score is the reconstruction error. The autoencoder gives similar results to the PCA\@. This is expected as autoencoders are a nonlinear extension of PCA\@.
    \item \textit{Multiple Objective Generative Adversarial Active Learning (MO-GAAL)}~\cite{GAAL}: One of the most advanced algorithms, using neural networks. It is composed of a discriminator and multiple generators. The generators try to imitate the data as best it can and the discriminator tries to distinguish the data, considered as inliers, from the generated events, considered as outliers.
\end{itemize}
PyOD also allows to use a combination of multiple algorithms.

\subsection{Signal Injection}
In order to choose the algorithm that will be the most sensitive, signal events are simulated. However, as stated previously it is not possible to have a representative simulation of all the possible signals, therefore this simulation is used for a proof of concept and to choose between multiple algorithms, but it does not allow us to get the sensitivity of the analysis to any signal.

The signal injection is done by picking a random direction and time and injecting signal in three or more of the datasets at this location accounting for the event's angular uncertainty and the detector's field-of-view.

\subsection{Output}
The outlier detection algorithm outputs an outlier score for each pixel of the skymap at each time step. However a signal is usually larger in extent than a single pixel and therefore adjacent pixels with a high outlier score are combined into one signal event. The outlier score of the event is the maximum score of its pixels.


\section{Status and Perspectives}
This analysis is approaching maturity, and we plan to use it on archives of five datasets AMON receives~\cite{AMON}: the ANTARES tracks, IceCube singlets, HAWC hotspots and HAWC bursts as well as Fermi LAT realtime data.

This analysis could be run in realtime in the future in order to send alerts and trigger follow-ups of the most significant outlier events. More datasets could also be added in the future without having to develop a new analysis.

\bibliographystyle{ICRC}
\bibliography{myBiblio}

%

%
%
%

\end{document}